\documentclass[preprint,showpacs,12pt,floatfix,aps]{revtex4}
 
\usepackage{graphicx}
\usepackage{latexsym}
\usepackage{amsmath}
\input psfig.sty

\newcommand{\bq}{\begin{equation}}
\newcommand{\eq}{\end{equation}}
\newcommand{\bqn}{\begin{eqnarray}}
\newcommand{\eqn}{\end{eqnarray}}
\newcommand{\nb}{\nonumber}
\newcommand{\lb}{\label}
\newcommand{\rr}{\bf r}
\begin{document}
\title{Gravastars with Dark Energy Evolving to Naked Singularity}
\author{R. Chan $^{1}$}
\email{chan@on.br}
\author{M.F.A. da Silva $^{2}$}
\email{mfasnic@gmail.com}
\author{P. Rocha $^{2,3}$}
\email{pedrosennarocha@gmail.com}
\author{J.F. Villas da Rocha $^{4}$}
\email{jfvroch@pq.cnpq.br}
\affiliation{\small $^{1}$ Coordena\c{c}\~ao de Astronomia e Astrof\'{\i}sica, 
Observat\'orio Nacional, Rua General Jos\'e Cristino, 77, S\~ao Crist\'ov\~ao,  
20921-400, Rio de Janeiro, RJ, Brazil\\
$^{2}$ Departamento de F\'{\i}sica Te\'orica, Instituto de F\'{\i}sica, 
Universidade do Estado do Rio de Janeiro, Rua S\~ao Francisco Xavier 524, 
Maracan\~a, 20550-900, Rio de Janeiro - RJ, Brasil\\
$^{3}$ IST - Instituto Superior de Tecnologia de Paracambi, FAETEC, Rua Sebasti\~ao de Lacerda
s/n, Bairro da F\'abrica, Paracambi, 26600-000, RJ, Brazil \\
$^{4}$ Universidade Federal do Estado do Rio de Janeiro,
Instituto de Bioci\^encias,
Departamento de Ci\^encias Naturais, Av. Pasteur 458, Urca,
22290-240, Rio de Janeiro, RJ, Brazil}
 
\date{\today}

\begin{abstract}
We consider a gravastar model made of anisotropic dark energy
with an infinitely thin spherical shell of a perfect fluid with the equation of
state $p = (1-\gamma)\sigma$
with an external de Sitter-Schwarzschild region.
It is found that in some cases the  models represent the "bounded excursion"
stable gravastars, where  the thin shell is oscillating between two finite radii,
while in other cases they collapse until the formation of black holes
or naked singularities.  An interesting result is that we can have
black hole and stable gravastar formation even with an interior and a shell constituted of dark
and repulsive dark energy, as also shown in previous work.
Besides, in three cases we have a dynamical evolution to a black hole (for $\Lambda=0$)
or to a naked singularity (for  $\Lambda > 0$). This is the first time in the literature that a naked
singularity emerges from a gravastar model.
\end{abstract}
  
\pacs{98.80.-k,04.20.Cv,04.70.Dy}

\maketitle

\section{Introduction}

As alternatives to black holes,  gravastars have received some attention
recently \cite{grava}, partially due to the tight connection between the 
cosmological constant and a currently accelerating universe \cite{DEs}, 
although very strict observational constraints on the existence of such 
stars may exist \cite{BN07}. 

The pioneer model of gravastar was proposed by Mazur and Mottola (MM) \cite{MM01},
consisting of five layers: an internal core
$0 < R < R_1$ , described by the de Sitter universe, an intermediate thin layer of stiff fluid
$R_1 < R < R_2$ , an external region $R > R_2$ , described by the Schwarzschild solution, and two 
infinitely thin shells, appearing, respectively, on the hypersurfaces $R = R_1$ and
$R = R_2$. The intermediate layer is constructed in such way that $R_1$ is inner than the de Sitter horizon, while $R_2$ is outer than the Schwarzschild horizon, eliminating the apparent horizon. Configurations with a de Sitter interior have long history which we can find, for example, in the work of Dymnikova and Galaktionov \cite{irina}.  
After this work, Visser and Wiltshire \cite{VW04} pointed out that there are 
two different types of stable gravastars which are stable gravastars and 
``bounded excursion" gravastars. In the spherically symmetric case, the motion 
of the surface of the gravastar can be written in the form \cite{VW04},
\bq
\lb{1.4}
\frac{1}{2}\dot{R}^{2} + V(R) = 0,
\eq
where $R$ denotes the radius of the star, and $\dot{R} \equiv dR/d\tau$, with 
$\tau$ being the proper time of the surface. Depending on the properties of 
the potential $V(R)$, the two kinds of gravastars are defined as follows. 

{\bf Stable gravastars}: In this case,  there must exist a radius $R_{0}$ such that
\bq
\lb{1.5}
V\left(R_{0}\right) = 0, \;\;\; V'\left(R_{0}\right) = 0, \;\;\;
V''\left(R_{0}\right) > 0,
\eq
where a prime denotes the ordinary differentiation with respect to the indicated argument.
If and only if there exists such a radius $R_{0}$ for which the above conditions are satisfied,
the model is said to be stable. 
Among other things, VW found that there are many equations of state for which the gravastar
configurations are stable, while others are not \cite{VW04}. Carter studied  the same
problem and found new equations of state for which the gravastar is stable \cite{Carter05}, 
while De Benedictis {\em et al} \cite{DeB06} and Chirenti and Rezzolla \cite{CR07} 
investigated the stability of the original model
of  Mazur and  Mottola against axial-perturbations, and found that gravastars are stable to
these perturbations, too. Chirenti and Rezzolla also showed that their quasi-normal modes 
differ from those of black holes with the same mass, and thus can be used to discern a gravastar 
from a black hole. 

{\bf "Bounded excursion" gravastars}: As VW noticed, there is a less stringent notion of 
stability, the so-called ``bounded excursion" models, in which there exist two radii $R_{1}$ 
and $R_{2}$ such that
\bq
\lb{1.6}
V\left(R_{1}\right) = 0, \;\;\; V'\left(R_{1}\right) \le 0, \;\;\;
V\left(R_{2}\right) = 0, \;\;\; V'\left(R_{2}\right) \ge 0,
\eq
with $V(R) < 0$ for $a \in \left(R_{1}, R_{2}\right)$, where $R_{2} > R_{1}$. 

Lately, we studied both types of gravastars \cite{JCAP,JCAP1}, and found that, 
such configurations can indeed be constructed, although   the region for the formation 
of them is very small in comparison to that of black holes.

Based on the discussions about the gravastar picture some authors have proposed
alternative models \cite{Chan}. Among them, we can find a Chaplygin dark star \cite{Paramos},
a gravastar supported by non-linear electrodynamics \cite{Lobo07},
a gravastar with continuous anisotropic pressure \cite{CattoenVisser05}
and recently, Dzhunushaliev et al. worked on spherically symmetric configurations
of a phantom scalar field and they found something like a gravastar but it was unstable\cite{singleton}.
In addition, Lobo \cite{Lobo} studied two models for a dark energy fluid. One of them
describes a homogeneous energy density and the other describes  an
ad-hoc monotonically decreasing energy density, although both of them are with anisotropic
pressure.  In order to match an exterior Schwarzschild spacetime he  
introduced a thin shell between the interior and the exterior spacetimes.

In this paper, we generalize our previous works \cite{JCAP,JCAP1} to the case where the 
equation of state of the infinitely thin shell is given by $p= (1-\gamma)  \sigma$ with 
$\gamma$ being a constant, the interior consists of a phantom energy fluid \cite{Lobo},
while the exterior is still the Schwarzschild space. 
We shall first construct three-layer dynamical models,  and then show both types of
gravastars and black holes exist for various situations. 
The rest of 
the paper is  organized as follows: In Sec. II we present the metrics of the 
interior and exterior spacetimes, and write down the motion of the thin shell
in the form of  equation (\ref{1.4}).  In Sec. III we show the definitions of dark and 
phantom energy, for the
interior fluid and for the shell.  In Sec. IV we discuss the formation of
black holes from standard or phantom energy. In Sec. V we analyze the formation of
gravastar or normal star from standard or phantom energy. In Sec. VI we
study special cases where we can not have the "bounded excursion".
Finally, in Sec. VII we present our conclusions.

\section{ Dynamical Three-layer Prototype Gravastars}

The interior fluid is made of an anisotropic dark energy fluid with a metric
given by \cite{Lobo}
\bq
ds^2_{-}=-f_1 dt^2 + f_2 dr^2 + r^2 d\Omega^2,
\lb{ds2-}
\eq
where  $d\Omega^2 \equiv d\theta^2 + \sin^2(\theta)d\phi^2$, and 
\bqn
f_1 &=& (1+b r^2)^{\frac{1-\omega}{2}}(1+2 b r^2)^{\omega},\nb\\
f_2 &=& \frac{1+2 b r^2}{1 + b r^2},
\eqn  
where $\omega$ is a constant, and its physical meaning can be seen from the
following equation (\ref{prpt}). Since the mass is given by 
$\bar m(r)=b r^3/[2(1+2br^2)]$ then we have that $b > 0$.
The corresponding energy density $\rho$, radial  and tangential  pressures $p_r$ and 
$p_t$ are given, respectively, by
\begin{eqnarray}
p_r&=&\omega \rho =\left(\frac{\omega
b}{8\pi}\right)\left(\frac{3+2b r^2}{(1+2b r^2)^2}\right), \nb \\
p_t&=& -\left(\frac{b}{8\pi}\right)\left(\frac{\omega(3+2b
r^2)}{(1+2b r^2)^2}\right)
+ \frac{b^2r^2}{32\pi\left[(1+2b r^2)^3(1+b r^2)\right]}\times \nb \\
&&\Big\{(1+\omega)(3+2b r^2)\left[(1+3\omega)+2br^2(1+\omega)\right] \nb \\
&&-8\omega(5+2br^2)(1+br^2)\Big\}.
\lb{prpt}
\end{eqnarray}
The exterior spacetime is given by the Schwarzschild metric
\bq
ds^2_{+}= - f dv^2 + f^{-1} d{\rr}^2 + {\rr}^2 d\Omega^2,
\lb{ds2+}
\eq
where $f=1 - {2m}/{\rr}-({\rr}/L_e)^2$ and $L_e=\sqrt{3/\Lambda_e}$.
The metric of the hypersurface  on the shell is given by
\bq
ds^2_{\Sigma}= -d\tau^2 + R^2(\tau) d\Omega^2.
\lb{ds2Sigma}
\eq
Since $ds^2_{-} = ds^2_{+} = ds^2_{\Sigma}$, we find that  $r_{\Sigma}={\rr}_{\Sigma}=R$,
and  
\bqn
\lb{dott2}
f_1\dot t^2 - f_2 \dot R^2 &=&  1,\\
\lb{dotv2}
f\dot v^2 - \frac{\dot R^2}{f} &=& 1,
\eqn
where the dot denotes the ordinary differentiation with respect to the proper time.
On the other hand,  the interior and exterior normal vectors to the thin shell are given by
\bqn
\lb{nalpha-}
n^{-}_{\alpha} &=& (-\dot R, \dot t, 0 , 0 ),\nb\\
n^{+}_{\alpha} &=& (-\dot R, \dot v, 0 , 0 ).
\eqn
Then, the interior and exterior extrinsic curvature are given by
\bqn
K^{i}_{\tau\tau}&=&\frac{1}{2} (1+b R^2)^{-\omega/2} \dot t \left\{ \left[ 4 (1+b R^2
)^{\omega/2} b R^2 \dot R^2+2 (1+b R^2)^{\omega/2} \dot R^2- \right. \right. \nb \\
& &\left. \left. (1+2 b R^2)^\omega \sqrt{1+b R^2} b R^2 \dot t^2-(1+2 b R^2)^ 
\omega \sqrt{1+b R^2} \dot t^2\right] (2 b R^2 \omega+2 b R^2+3 \omega+1)- \right. \nb\\
& &\left. 2 (1+b R^ 2)^{\omega/2} (1+2 b R^2) \dot R^2 \right\} (1+2 b R^2
)^{-2} (1+b R^2)^{-1} b R+ \dot R \ddot t- \ddot R \dot t,\\
\lb{Ktautau-}
\eqn
\bq
K^{i}_{\theta\theta} = \frac{\dot t(1+b R^2) R}{1 + 2 b R^2},
\lb{Kthetatheta-}
\eq
\bq
K^{i}_{\phi\phi} = K^{-}_{\theta\theta}\sin^2(\theta),
\lb{Kphiphi-}
\eq
\bqn
K^{e}_{\tau\tau}&=&\dot v [(2 L_e^2 m \dot v+L_e^2 R \dot R-L_e^2 R \dot v+R^
3 \dot v) (2 L_e^2 m \dot v-L_e^2 R \dot R-L_e^2 R \dot v+R^3 \dot v)- \nb \\
& &2 L_e^4 R^2 \dot R^2] ((2 m-R) L_e^2+R^ 3)^{-1}
(L_e^2 m-R^3) L_e^{-4} R^{-3}+\dot R \ddot v- \ddot R \dot v
\lb{Ktautau+}
\eqn
\bq
K^{e}_{\theta\theta}= -\dot v((2 m-R) L_e^2+R^3) L_e^{-2}
\lb{Kthetatheta+}
\eq
\bq
K^{e}_{\phi\phi}=K^{e}_{\theta\theta}\sin^2(\theta).
\lb{Kphiphi+}
\eq

Since \cite{Lake}
\bq
[K_{\theta\theta}]= K^{e}_{\theta\theta}-K^{i}_{\theta\theta} = - M,
\lb{M}
\eq
where $M$ is the mass of the shell, we find that
\bq
M=\dot v (2 m-R)+\frac{\dot t(1+b R^2) R}{1 + 2 b R^2}.
\lb{M1}
\eq
Then, substituting equations (\ref{dott2}) and (\ref{dotv2}) into (\ref{M1}) 
we get
\bq
M=-R\left(1-\frac{2m}{R} -\left(\frac{R}{L_e}\right)^2 + \dot R^2 \right)^{1/2} + 
R\frac{ \left[ 1 + b R^2 + \dot R^2 (1 + 2 b R^2) \right]^{1/2}}
{(1+b R^2)^{-(\omega+1)/4}(1+2b R^2)^{(\omega+2)/2}}.
\lb{M2}
\eq
In order to keep the ideas of MM as much as possible, we consider the thin 
shell as consisting
of a fluid with the equation of state, $p=(1-\gamma)\sigma$, where $\sigma$ and $p$ denote, 
respectively, the surface energy density and pressure of the shell and $\gamma$ is a constant. 
Then, the equation of motion of the shell is given by \cite{Lake}
\bq
\dot M + 8\pi R \dot R p = 4 \pi R^2 [T_{\alpha\beta}u^{\alpha}n^{\beta}]=
4\pi R^2 \left(T^e_{\alpha\beta}u_e^{\alpha}n_e^{\beta}-T^i_{\alpha\beta}u_i^{\alpha}n_i^{\beta} \right),
\lb{dotM}
\eq
where $u^{\alpha}$ is the four-velocity.  Since the interior fluid is made
of an anisotropic fluid and the exterior is vacuum, we get
\bq
\dot M + 8\pi R \dot R (1-\gamma)\sigma = 0.
\lb{dotM1}
\eq
Recall that $\sigma = M/(4\pi R^2)$, we find that  equation (\ref{dotM1}) has the solution
\bq
M=k R^{2(\gamma-1)},
\lb{Mk}
\eq
where $k$ is an integration constant. Substituting  equation (\ref{Mk}) into  equation (\ref{M2}),
and rescaling $m, \; b$ and $R$ as,
\bqn
m &\rightarrow& mk^{-\frac{1}{2\gamma-3}},\nb\\
b &\rightarrow& b k^{\frac{2}{2\gamma-3}},\nb\\
R &\rightarrow& Rk^{-\frac{1}{2\gamma-3}},
\eqn
we find that it can be written in the form of equation (\ref{1.4}), and
\bqn
V(R,m,L_e,\omega,b,\gamma)&=& -\frac{1}{2L_e^2 R^2 b_2 \left( b_2^{\omega+1} b_1^{-\frac{1}{2} \omega}-b_1^{\frac{1}{2}} \right)^2} (b_2^{2 \omega+3} R^{4 \gamma-4} b_1^{-\omega} L_e^2 \nb \\
& &-2 b_1^{-\frac{1}{2} \omega+\frac{1}{2}} R^{2 \gamma-2} b_2^{\omega+1} L_e (-b_2 (b_2^{\omega+2} b_1^{-\frac{1}{2} \omega} R^6 L_e^2-2 b_2^{\omega+2} b_1^{-\frac{1}{2} \omega} m L_e^2 R^5- \nb \\
& &b_2^{\omega+2} b_1^{-\frac{1}{2} \omega} R^8-b_2^{\omega+1} b_1^{-\frac{1}{2} \omega+1} R^6 L_e^2-R^6 L_e^2 b_2 b_1^{\frac{1}{2}}+2 m L_e^2 b_2 R^5 b_1^{\frac{1}{2}}+ \nb \\
& &R^8 b_2 b_1^{\frac{1}{2}}+R^6 L_e^2 b_1^{\frac{3}{2}}-b_2^{\omega+2} L_e^2 b_1^{-\frac{1}{2} \omega} R^{4 \gamma})/(b_1^{1/2} R^4))^{\frac{1}{2}}+b_1^{\frac{3}{2}-\frac{1}{2} \omega} b_2^{\omega+1} R^2 L_e^2- \nb \\
& &b_1^2 R^2 L_e^2+b_2^{\omega+2} b_1^{-\frac{1}{2} \omega+\frac{1}{2}} L_e^2 R^{4 \gamma-4}-b_2^{2 \omega+3} L_e^2 b_1^{-\omega} R^2+ b_2^{\omega+2} b_1^{-\frac{1}{2} \omega+\frac{1}{2}} R^2 L_e^2+ \nb \\
& &2 b_2^{2 \omega+3} b_1^{-\omega} R m L_e^2-2 b_2^{\omega+2} b_1^{-\frac{1}{2} \omega+\frac{1}{2}} m L_e^2 R+b_2^{2 \omega+3} b_1^{-\omega} R^4-b_2^{\omega+2} b_1^{-\frac{1}{2} \omega+\frac{1}{2}} R^4) \nb \\
\lb{VR}
\eqn
where
\bq
\lb{b1}
b_1 \equiv 1+b R^2,\;\;\; 
b_2 \equiv 1+2 b R^2.
\eq

The exterior horizons are given by \cite{Shankaranarayanan}
\bq
r_{bh}= \frac{2m}{\sqrt{3 y}} \cos \left( \frac{\pi+\psi}{3} \right),
\label{rbh}
\eq
\bq
r_c= \frac{2m}{\sqrt{3 y}} \cos \left( \frac{\pi-\psi}{3} \right),
\label{rc}
\eq
where $y=(m/L_e)^2$, $\psi= \arccos \left( 3\sqrt{3 y} \right)$, $r_{bh}$ denotes the black
hole horizon and $r_c$ denotes the cosmological horizon.  
Note that if $y > 1/27$
the quantity $3\sqrt{3 y}$ is greater than 1, giving an imaginary angle $\psi$.
Thus, the
horizons $r_{bh}$ and $r_c$ are imaginary and the spacetime becomes free of
horizons.

To  guarantee
that initially the spacetime does not have any kind of horizons,  cosmological or event,
we must restrict $R_{0}$ to the range,
\bq
\lb{2.2b}
r_{bh} < R_{0} < r_h \; or \; r_c,
\eq
where $R_0$ is the initial collapse radius.
Clearly, for any given constants $m$, $\omega$, $b$ and $\gamma$, equation (\ref{VR}) uniquely 
determines the collapse of the prototype  gravastar. Depending on the initial value $R_{0}$,  
the collapse can form either a black hole,  a gravastar,   a Minkowski, or a spacetime filled with
phantom fluid. In the last case, the thin shell
first collapses to a finite non-zero minimal radius and then expands to infinity.  To  guarantee
that initially the spacetime does not have any kind of horizons,  cosmological or event,
we must restrict $R_{0}$ to the range,
\bq
\lb{2.2c}
R_{0} > 2 m,
\eq
where $R_0$ is the initial collapse radius. When $m = 0= b$, the thin shell disappears,
and the whole spacetime is Minkowski. So, in the following we shall not consider this case.

Since the potential  (\ref{VR}) is so complicated, it is too difficult to study it
analytically. Instead, in the following we shall study it numerically. Before doing so, we
shall show the classifications of matter, dark energy, and phantom energy for 
anisotropic fluids.

\section{Classifications of Matter, Dark Energy, and Phantom Energy  for Anisotropic Fluids}

Recently \cite{Chan08}, the classification of matter, dark and phantom energy 
for an anisotropic fluid was given  in terms of the energy conditions. Such a classification
is necessary for systems where anisotropy is important, and  the pressure components 
may play very important roles and  can have quite different contributions.
In this paper, we will use this classification to study the collapse of the
dynamical prototype gravastars, constructed in the last section. 
In particular, we define dark
energy  as a fluid which violates the strong energy
condition
(SEC).  From the Raychaudhuri equation, we can see that such defined
dark energy always exerts  divergent forces on  time-like or null geodesics.
On the other hand,  we define phantom energy as a fluid  that violates at least
one of the null energy conditions (NEC's). We shall further distinguish phantom
energy that satisfies the SEC
from that which does not satisfy the SEC. We call
the former attractive 
phantom energy, and the latter
 repulsive phantom energy.
Such a classification is summarized in Table I.

For the sake of completeness, in Table II we apply it to the matter field
located on the thin shell, while in Table III we combine all the results of Tables I 
and II, and present all the possibilities.   

\begin{table}
\caption{\label{tab:table1} This table summarizes the  classification of
the interior matter field, based on the energy conditions \cite{HE73}, where
we assume that $\rho \ge 0$.}
\begin{ruledtabular}
\begin{tabular}{cccc}
Matter & Condition 1 & Condition 2  & Condition 3 \\
\hline
Normal Matter           & $\rho+p_r+2p_t\ge 0$ & $\rho+p_r\ge 0$ & $\rho+p_t\ge 0$ \\
\hline
Dark Energy               & $\rho+p_r+2p_t <  0$ & $\rho+p_r\ge 0$ & $\rho+p_t\ge 0$ \\
\hline
                          &                      & $\rho+p_r <  0$ & $\rho+p_t\ge 0$ \\
Repulsive Phantom Energy  & $\rho+p_r+2p_t <  0$ & $\rho+p_r\ge 0$ & $\rho+p_t <  0$ \\
                          &                      & $\rho+p_r <  0$ & $\rho+p_t <  0$ \\
\hline
Attractive Phantom Energy & $\rho+p_r+2p_t\ge 0$ & $\rho+p_r\ge 0$ & $\rho+p_t <  0$ \\
                          &                      & $\rho+p_r <  0$ & $\rho+p_t\ge 0$ \\
\end{tabular}
\end{ruledtabular}
\end{table}

\begin{table}
\caption{\label{tab:table2} This table summarizes the  classification of matter
on the thin shell, based on the energy conditions \cite{HE73}. The last column indicates
the particular values of the parameter $\gamma$, where we assume that $\rho \ge 0$.}
\begin{ruledtabular}
\begin{tabular}{cccc}
Matter & Condition 1 & Condition 2  & $\gamma$ \\
\hline
Normal Matter           & $\sigma+2p\ge 0$ & $\sigma+p\ge 0$ & -1 or 0  \\
Dark Energy               & $\sigma+2p <  0$ & $\sigma+p\ge 0$ &  7/4 \\
Repulsive Phantom Energy  & $\sigma+2p <  0$ & $\sigma+p <  0$ &   3  \\
\end{tabular}
\end{ruledtabular}
\end{table}

In order to consider the equations (\ref{ds2-}) and (\ref{prpt}) for describing dark energy
stars we must analyze carefully the ranges of the parameter $\omega$ that in
fact furnish the expected fluids.  It can be shown that the
condition $\rho+p_r>0$ is violated for $\omega<-1$ and fulfilled for $\omega>-1$, 
for any values of $R$ and $b$.
The conditions $\rho+p_t>0$ and $\rho+p_r+2p_t>0$ are satisfied for $\omega<-1$
and $-1/3<\omega<0$, for any values of $R$ and $b$.  
For the other intervals of
$\omega$  the
 energy conditions depend
on very complicated relations of $R$ and $b$.  See reference \cite{Chan08}. 
This provides an explicit example, in which the definition of dark energy must 
be dealed with great care.  Another case was provided in a previous work \cite{Chan08}.  
Taking several values of $\omega$ in the intervals $-1<\omega<-1/3$ and 
$\omega>0$, we could not found any case where the interior dark energy exist.

In order to fulfill the energy condition $\sigma+2p\ge0$ of the shell
and assuming that
$p=(1-\gamma)\sigma$ we must have $\gamma \le 3/2$. On the other hand, in order
to satisfy the condition $\sigma+p\ge 0$, we obtain $\gamma \le 2$.
Hereinafter, we will use only some particular values of the parameter
$\gamma$ which are analyzed in this work. See Table II.

\section{Structures Formed}

Here we can find many types of systems, depending on the combination of the 
constitution matter of the shell and core.  Among them, there are formation of
black holes, stable and "bounded excursion" gravastars, as it has already 
shown in our previous works \cite{JCAP}-\cite{JCAP4}, 
and even a naked singularity constituted exclusively of dark energy. 
All of them are  listed in the table III. 

As can be seen in the figures \ref{fig1}, \ref{fig3} and \ref{fig5}, depending 
on the value of the cosmological constant, we can see that $V(R) = 0$ now can i
have one, two  or three real roots. Then, we 
have, say, $R_{i}$, where $R_{i+1} > R_{i}$. For $L_e=L_1$ (corresponding to 
$\Lambda=0$)  If we choose $R_{0} > R_{3}$ none structure is allowed in this 
region because the potential is greater than the zero.  However, if we choose 
$R_{2} < R_{0} < R_{3}$, the collapse will bounce back and forth between 
$R = R_{1}$ and $R = R_{2}$. This is exactly the so-called "bounded excursion" 
model mentioned in \cite{VW04}, and studied in some details in 
\cite{JCAP}-\cite{JCAP4}.  Of 
course, in a realistic situation, the star will emit both gravitational waves 
and particles, and the potential will be self-adjusted to produce a minimum at 
$R = R_{static}$ where $V\left(R=R_{static}\right) = 0 = V'\left(R=R_{static}\right)$ 
whereby a gravastar is finally formed \cite{VW04,JCAP,JCAP1,JCAP2}. 
For $R_{0} < R_{1}$ a black hole is formed in the end of the collapse of the shell. 

The scenario above can significantly be changed if we consider $\Lambda>0$. 
In this case for $L_e>L_c$, we also have bounded excursion gravastars if 
$R_{2} < R_{0} < R_{3}$. However, for $R_{0} < R_{1}$ the final structure 
can be now a black hole or a naked singularity since the presence of the 
cosmological constant above a certain limit ($L_e^*$) eliminates the 
event horizon (its radius becomes imaginary), as can be seen in the tables IV,
V and VI.  This is the first 
evidence of a naked singularity formation from a gravastar model. Moreover 
for $L_e=L_c$, then $R_{2} = R_{3}$, a stable gravastar is formed if 
$R_{0} = R_{2}$, while for $L_e<L_c$ there is only one real root. Note that 
for any value of $L_e>L_e^*$, a naked singularity is formed for small 
initial radius of the shell.
This is already present in the de Sitter-Schwarzschild solution
\cite{Shankaranarayanan}, since is the exterior cosmological constant which allows
to relax the inevitability of the horizon formation (differently from the
Schwarzschild solution). The news here are that we have found a source (the shell
with an interior anisotropic dark energy fluid) which can be matched with the de 
Sitter-Schwarzschild vacuum
spacetime and, depending on the values of the exterior cosmological
constant and the total mass, can represent a naked singularity.

Thus, solving equation (\ref{M2}) for $\dot R(\tau)$ we can integrate 
$\dot R(\tau)$ and obtain $R(\tau)$, which are
shown in the figures \ref{fig2}, \ref{fig4} and \ref{fig6} for the case G.
 
\begin{table}
\caption{\label{tab:table3}This table summarizes all possible kind of energy
of the interior fluid and of the shell and compares the formed structures
in the two gravastar models ($\Lambda_e=0$, $\Lambda_e>0$). The letters SG, UG, BEG, BH, NS and N
denote stable gravastar, unstable gravastar, bounded excursion gravastar,
black hole, naked singularity and none, respectively.}
\begin{ruledtabular}
\begin{tabular}{cccccc}
Case & Interior Energy & Shell Energy & Figures & Structures ($\Lambda_e=0$) & Structures ($\Lambda_e>0$)\\
\hline
A & Standard           & Standard           &   & SG     & SG/UG/BEG \\
B & Standard           & Dark               &   & BH     & SG/UG/BEG \\
C & Standard           & Repulsive Phantom  &   & BH     & BH \\
D & Dark               & Standard           &   & N      & N \\
E & Dark               & Dark               &   & N      & N \\
F & Dark               & Repulsive Phantom  &   & N      & N \\
G & Repulsive Phantom  & Standard           & \ref{fig1}, \ref{fig3}, \ref{fig5}  & SG/BEG & BH/SG/UG/BEG/NS \\
H & Repulsive Phantom  & Dark               &   & BH     & BH/SG/UG/BEG \\
I & Repulsive Phantom  & Repulsive Phantom  &   & BH     & BH \\
J & Attractive Phantom & Standard           &   & SG/BEG & BH/SG/BEG \\
K & Attractive Phantom & Dark               &   & BH     & BH \\
L & Attractive Phantom & Repulsive Phantom  &   & BH     & BH \\
\end{tabular}
\end{ruledtabular}
\end{table}

\section{Conclusions}

In this paper, we have studied the problem of the stability of gravastars by
constructing dynamical three-layer models  of VW \cite{VW04},
which consists of an internal anisotropic dark energy fluid, a dynamical 
infinitely thin  shell of perfect fluid with the equation of state 
$p = (1-\gamma)\sigma$, and an external de Sitter-Schwarzschild spacetime.

We have shown explicitly that the final output can be a black
hole, a "bounded excursion" stable gravastar depending on the total mass
$m$ of the system, the cosmological constant $L_e$, the parameter $\omega$,
the constant $b$, the parameter $\gamma$ and
the initial position $R_{0}$ of the dynamical shell. All these possibilities
have non-zero measurements in the phase space of $m$, $L_e$, $b$, $\omega$, 
$\gamma$ and $R_{0}$.  All the results can be summarized in Table III.

\begin{figure}
\vspace{.2in}
\centerline{\psfig{figure=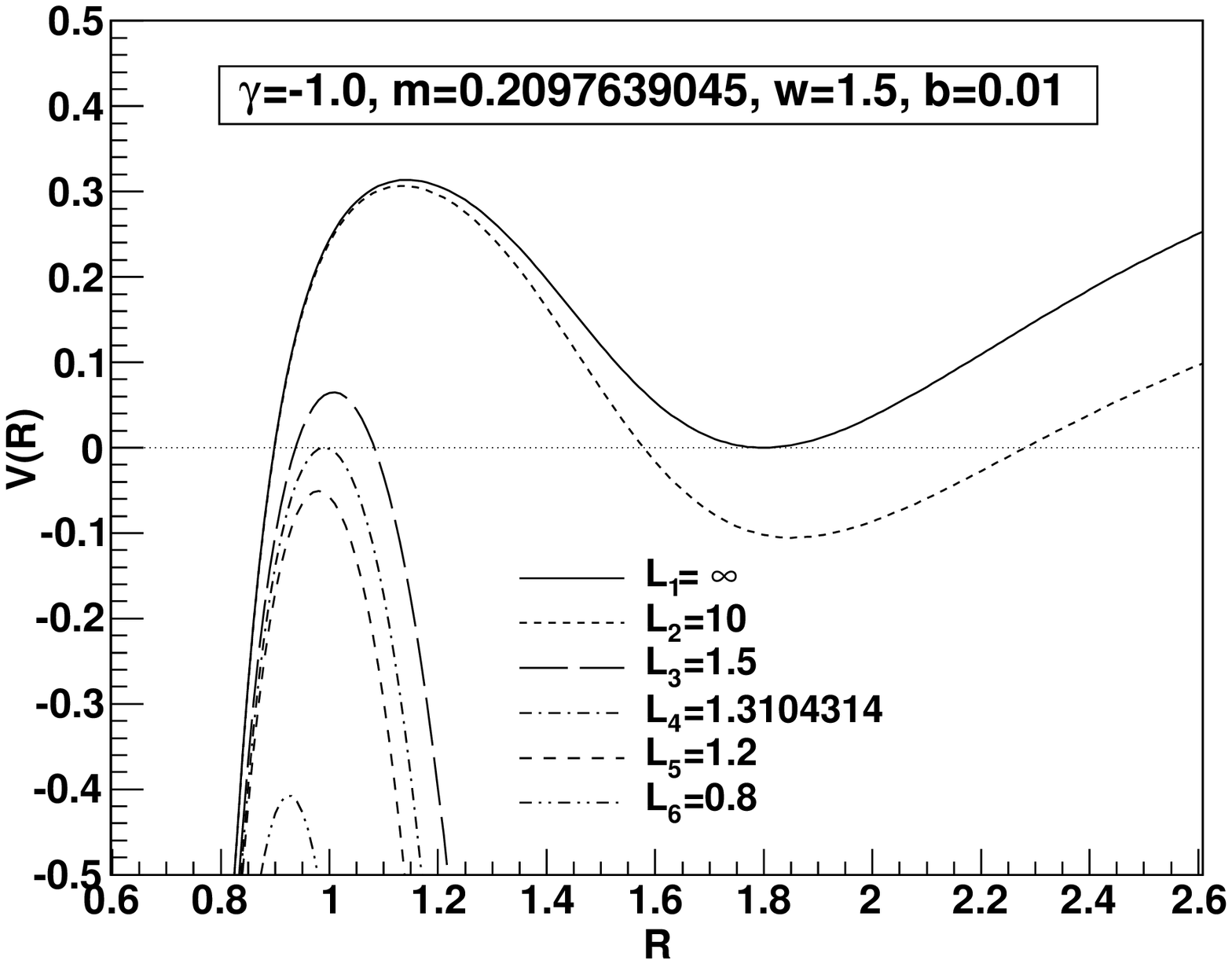,width=3.3truein,height=3.0truein}\hskip
.25in \psfig{figure=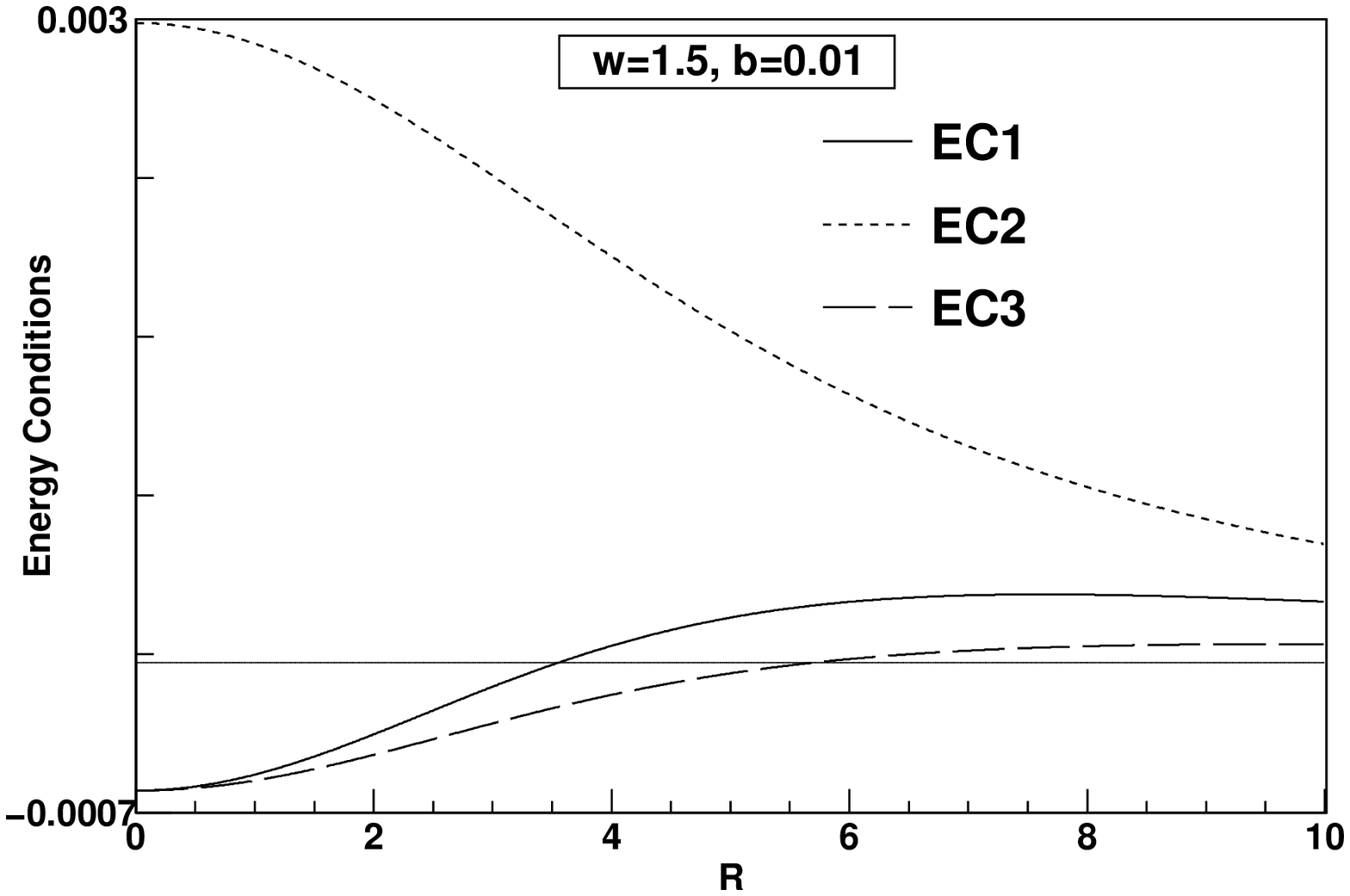,width=3.3truein,height=3.0truein}
\hskip .5in} \caption{The potential $V(R)$ and the energy conditions EC1$\equiv \rho+p_r+2p_t$, 
EC2$\equiv \rho+p_r$ and EC3$\equiv \rho+p_t$, for $\gamma=-1$,
$\omega=1.5$, $b=0.01$ and $m_c=0.2097639045$. {\bf Case G}}
\label{fig1}
\end{figure}

\begin{figure}
\vspace{.2in}
\centerline{\psfig{figure=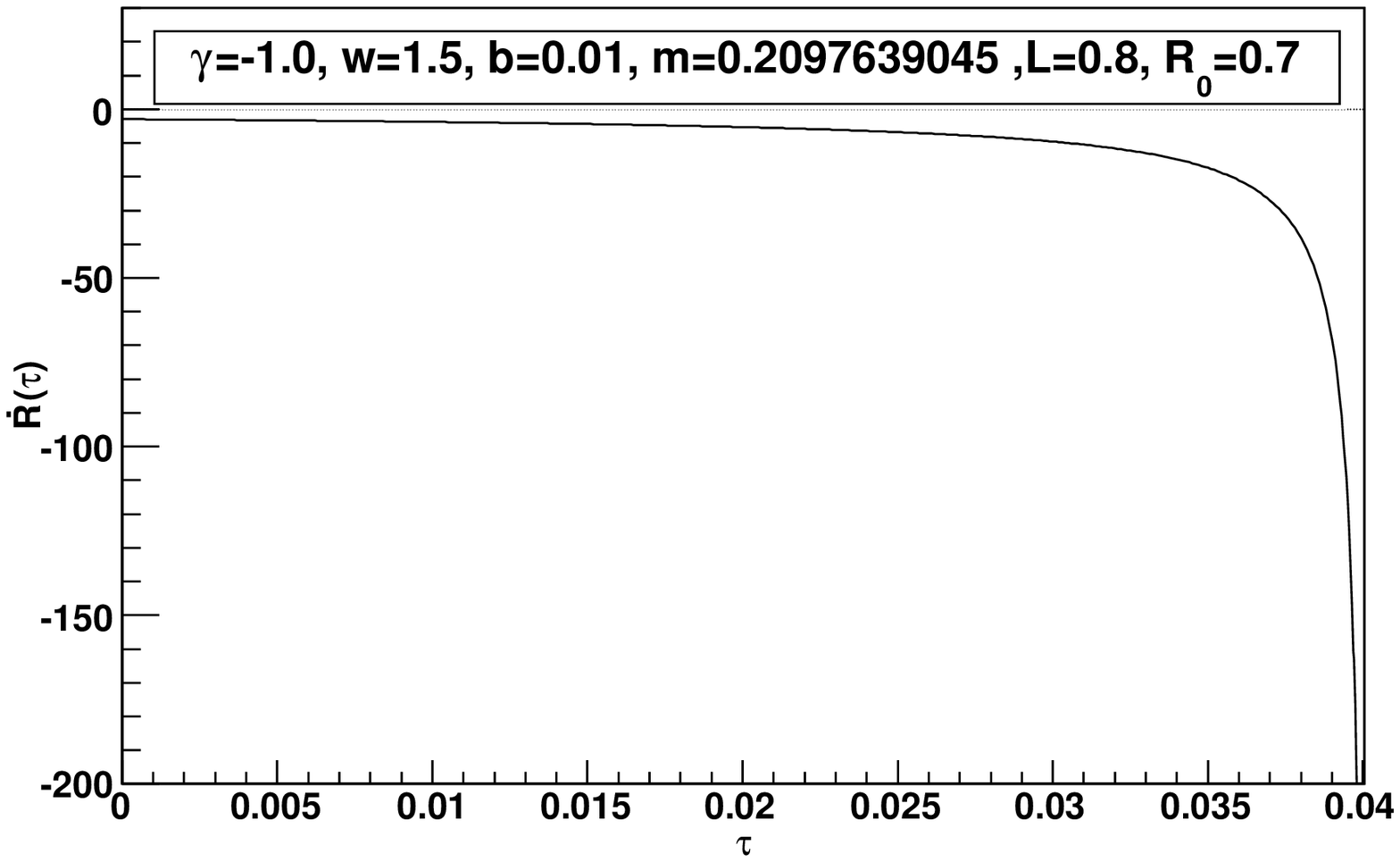,width=3.3truein,height=3.0truein}\hskip
.25in \psfig{figure=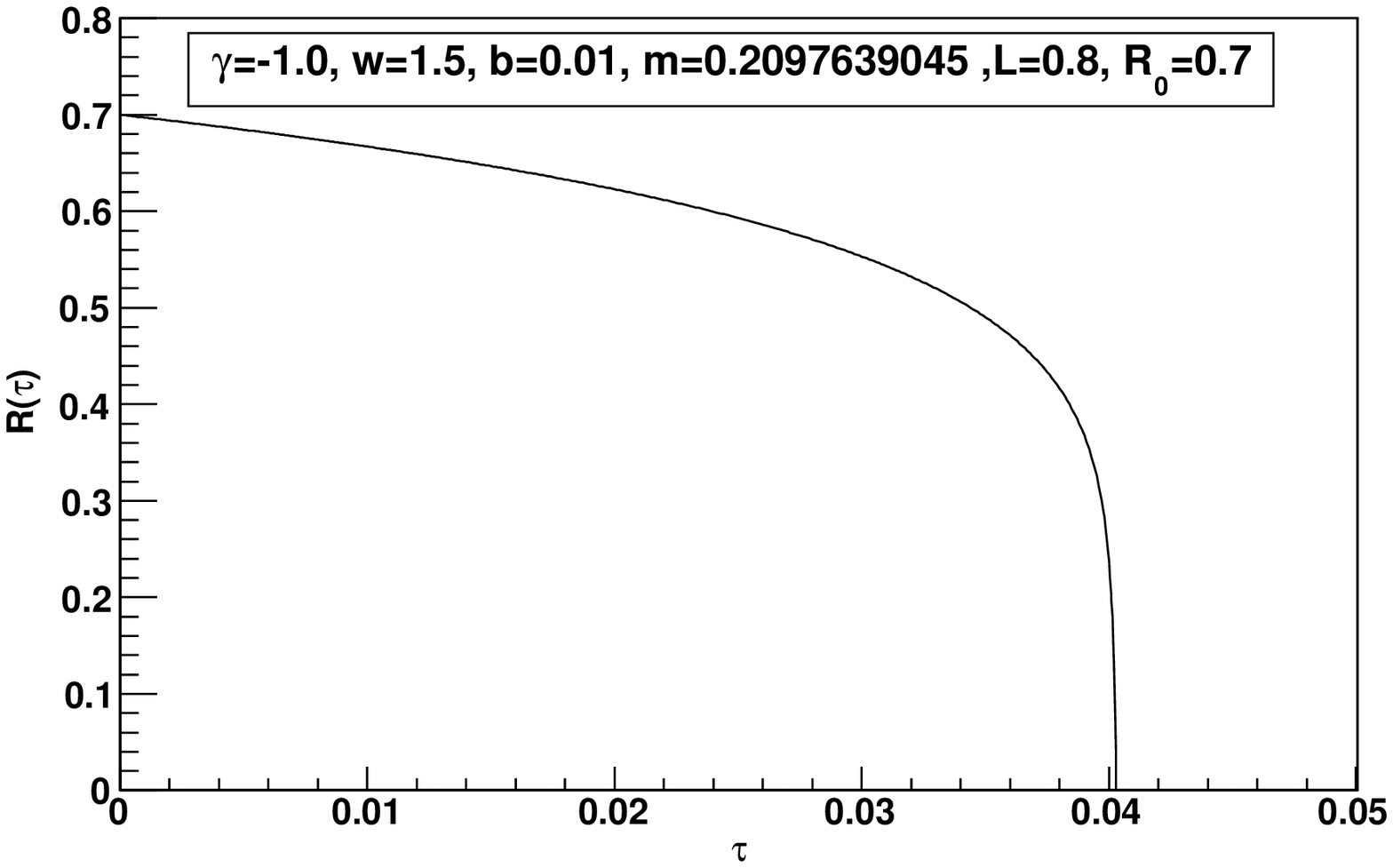,width=3.3truein,height=3.0truein}
\hskip .5in} \caption{These figures represent the dynamical evolution of the shell to a 
naked singularity for the potential given by the figure \ref{fig1},
assuming $R_0=R(0)=0.7$.}
\label{fig2}
\end{figure}

\begin{figure}
\vspace{.2in}
\centerline{\psfig{figure=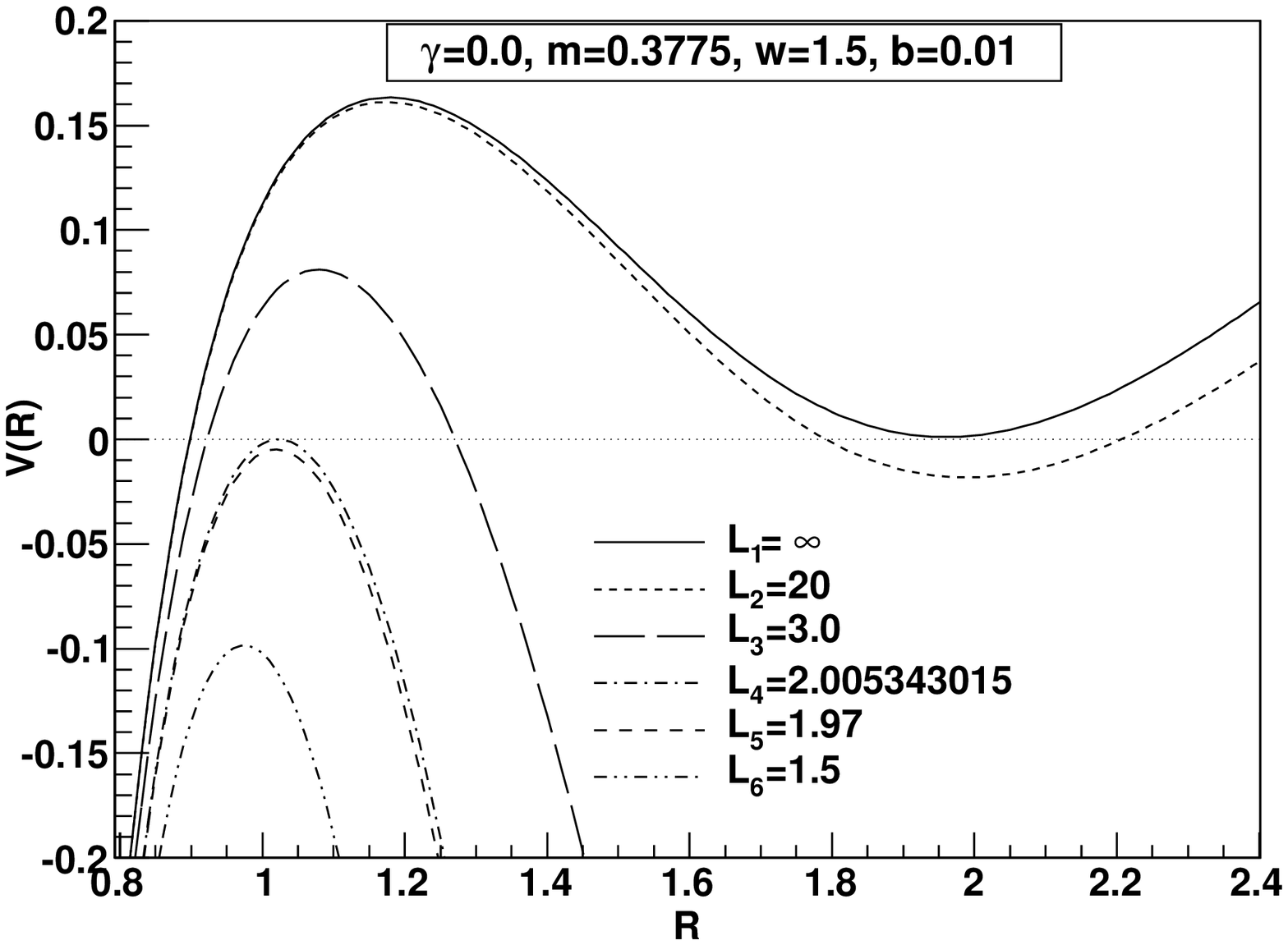,width=3.3truein,height=3.0truein}\hskip
.25in \psfig{figure=ECw1v5b0v01.eps,width=3.3truein,height=3.0truein}
\hskip .5in} \caption{The potential $V(R)$ and the energy conditions EC1$\equiv \rho+p_r+2p_t$, 
EC2$\equiv \rho+p_r$ and EC3$\equiv \rho+p_t$, for $\gamma=0$,
$\omega=1.5$, $b=0.01$ and $m_c=0.3775$. {\bf Case G}}
\label{fig3}
\end{figure}

\begin{figure}
\vspace{.2in}
\centerline{\psfig{figure=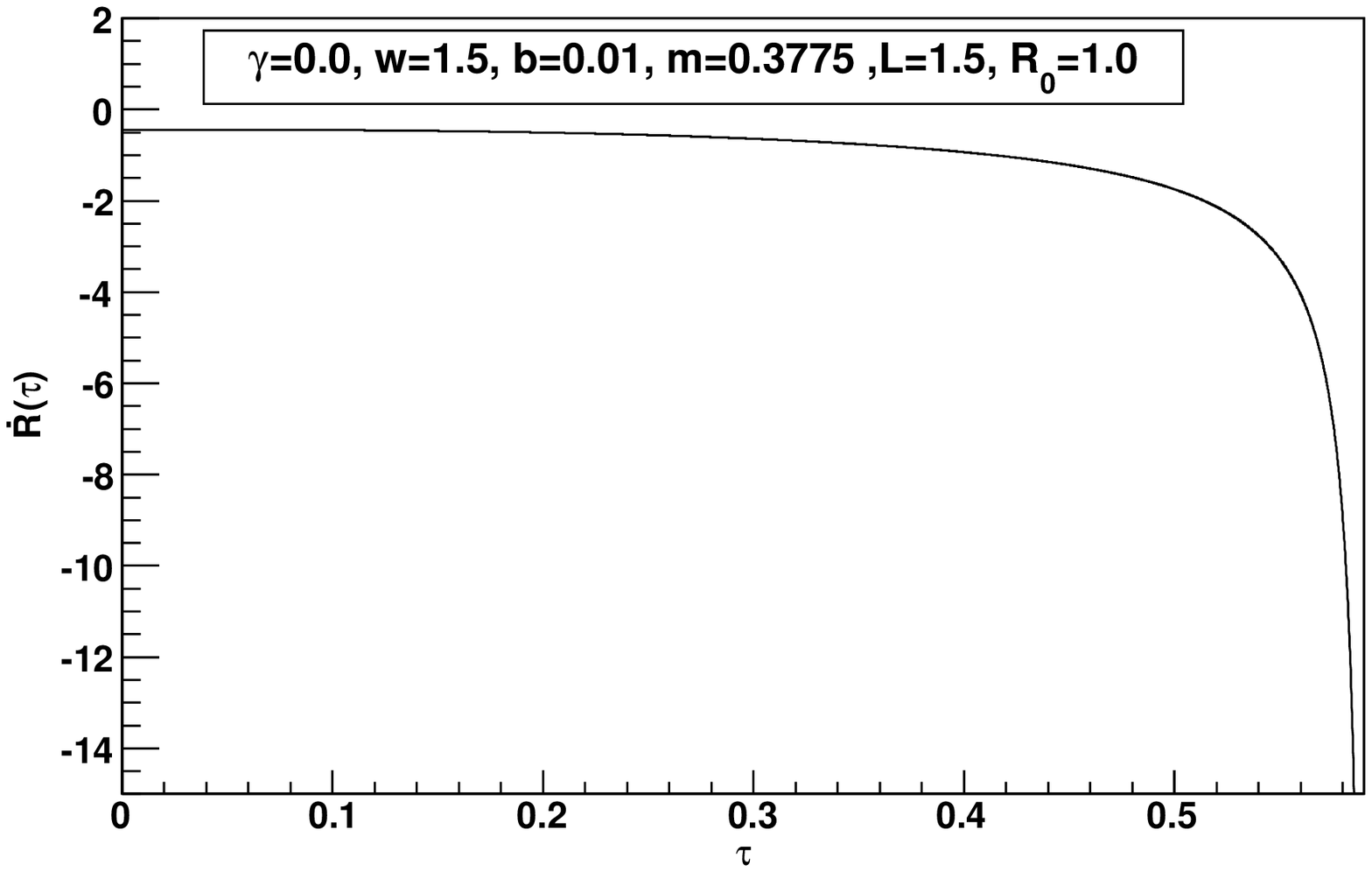,width=3.3truein,height=3.0truein}\hskip
.25in \psfig{figure=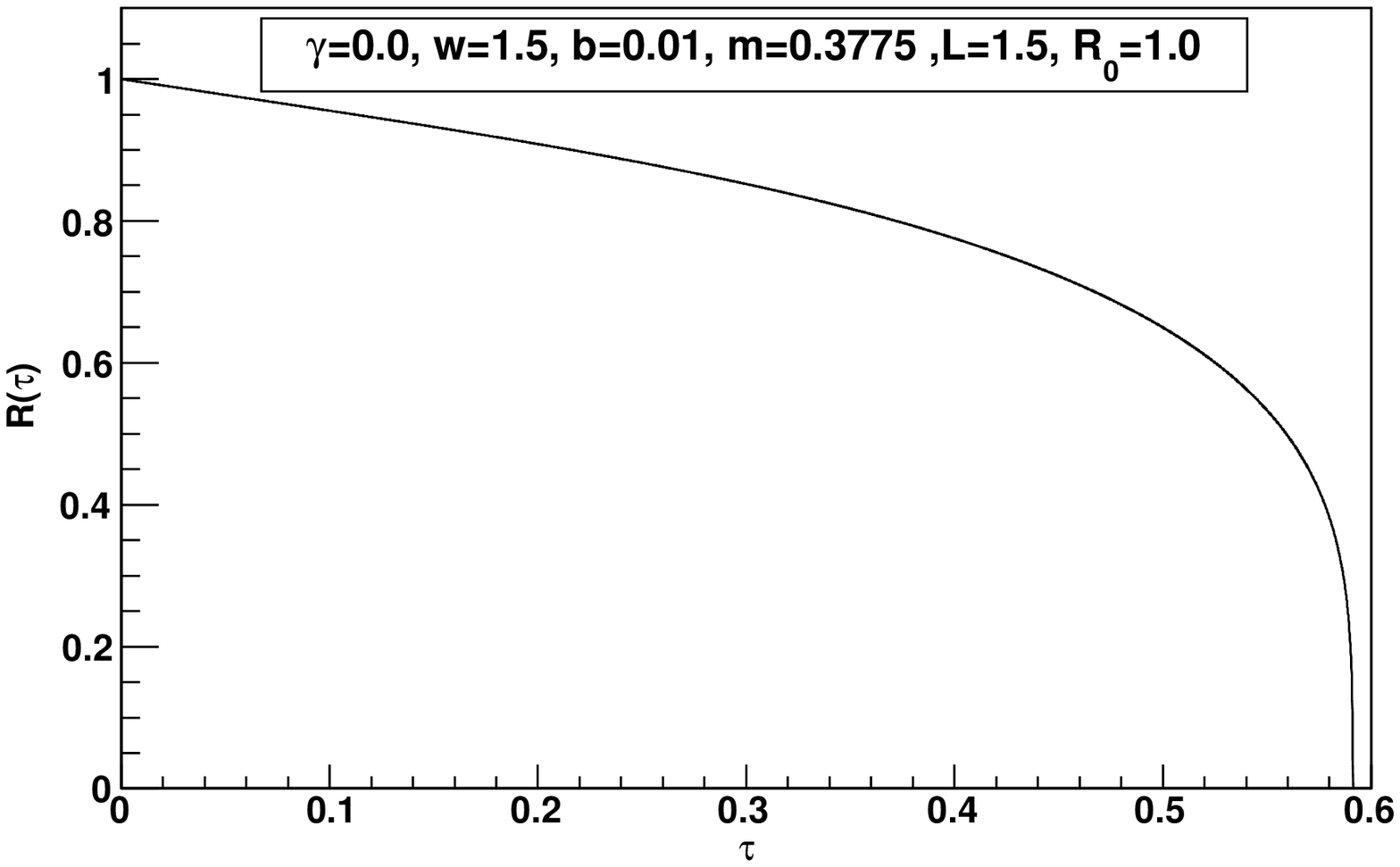,width=3.3truein,height=3.0truein}
\hskip .5in} \caption{These figures represent the dynamical evolution of the shell to a 
naked singularity for the potential given by the figure \ref{fig3},
assuming $R_0=R(0)=1$.}
\label{fig4}
\end{figure}

\begin{figure}
\vspace{.2in}
\centerline{\psfig{figure=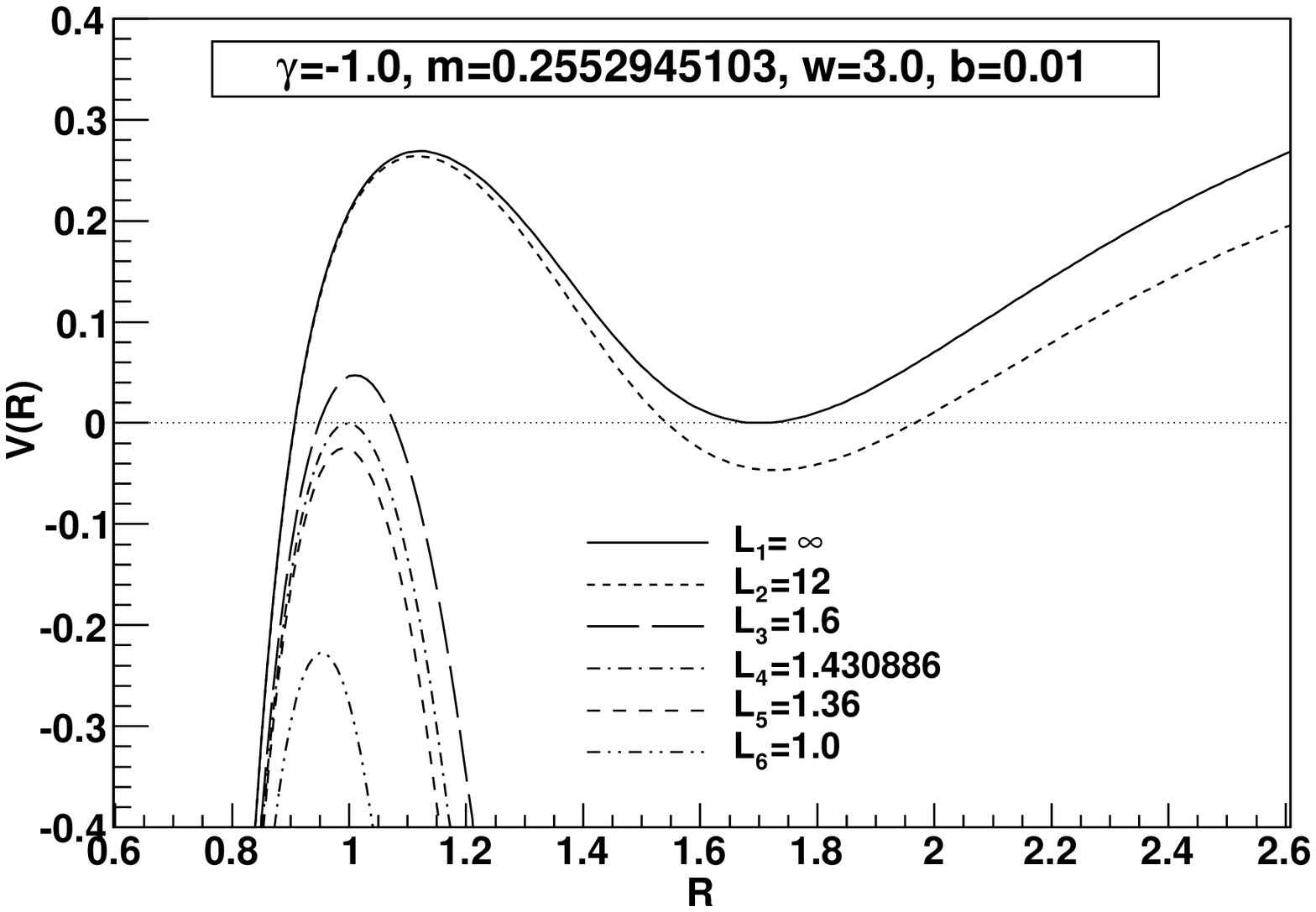,width=3.3truein,height=3.0truein}\hskip
.25in \psfig{figure=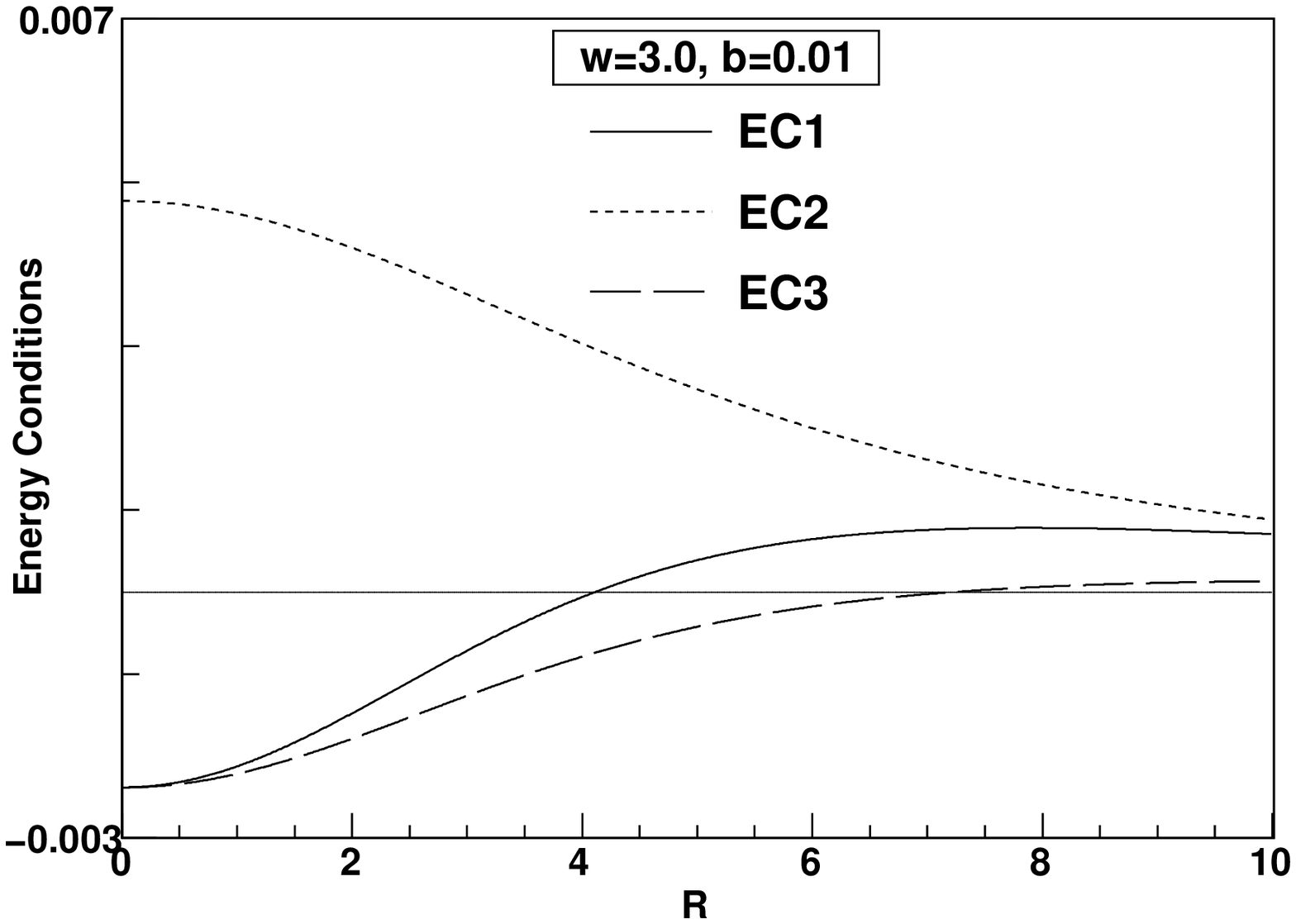,width=3.3truein,height=3.0truein}
\hskip .5in} \caption{The potential $V(R)$ and the energy conditions EC1$\equiv \rho+p_r+2p_t$, 
EC2$\equiv \rho+p_r$ and EC3$\equiv \rho+p_t$, for $\gamma=-1$,
$\omega=3$, $b=0.01$ and $m_c=0.2552945103$. {\bf Case G}}
\label{fig5}
\end{figure}

\begin{figure}
\vspace{.2in}
\centerline{\psfig{figure=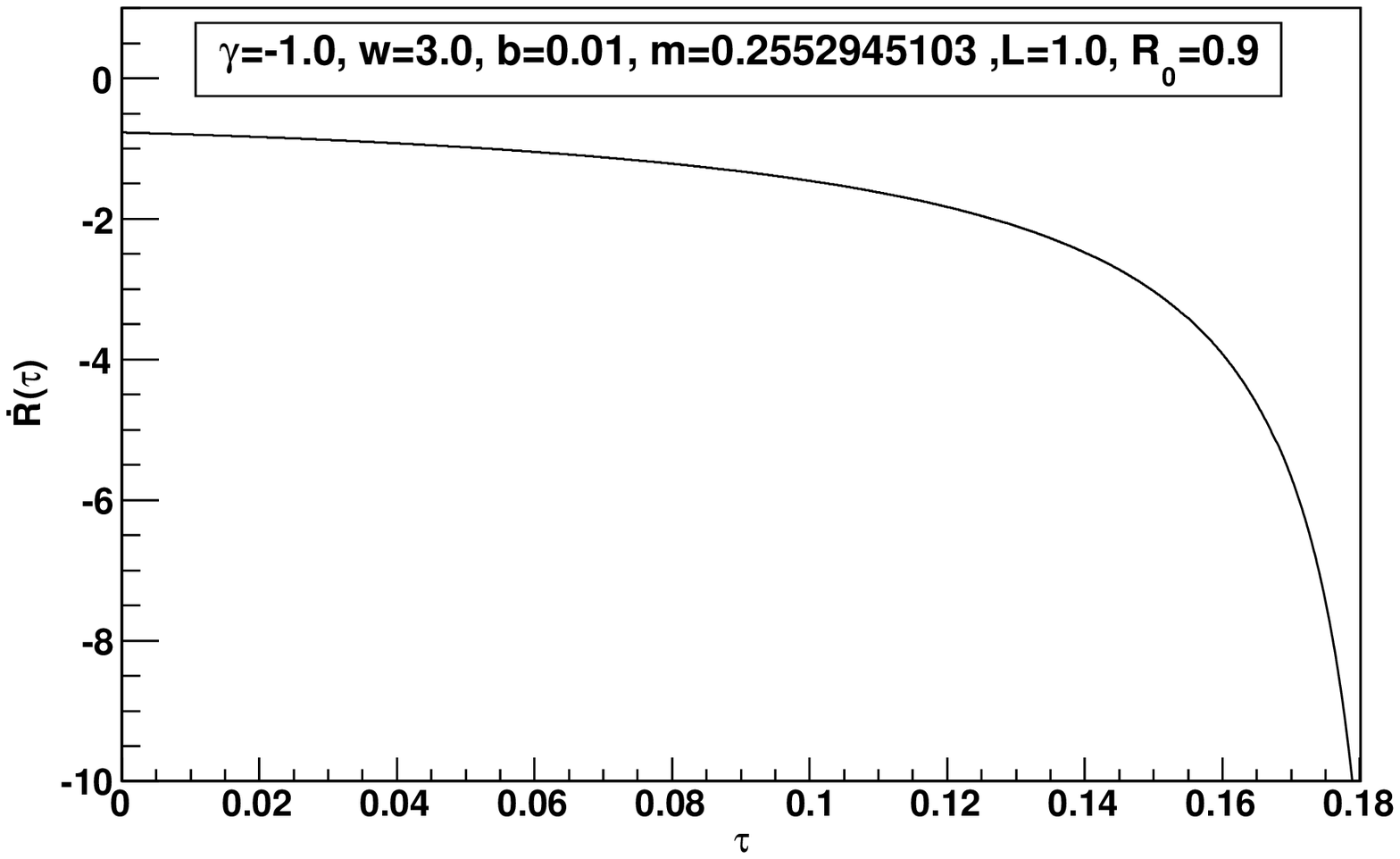,width=3.3truein,height=3.0truein}\hskip
.25in \psfig{figure=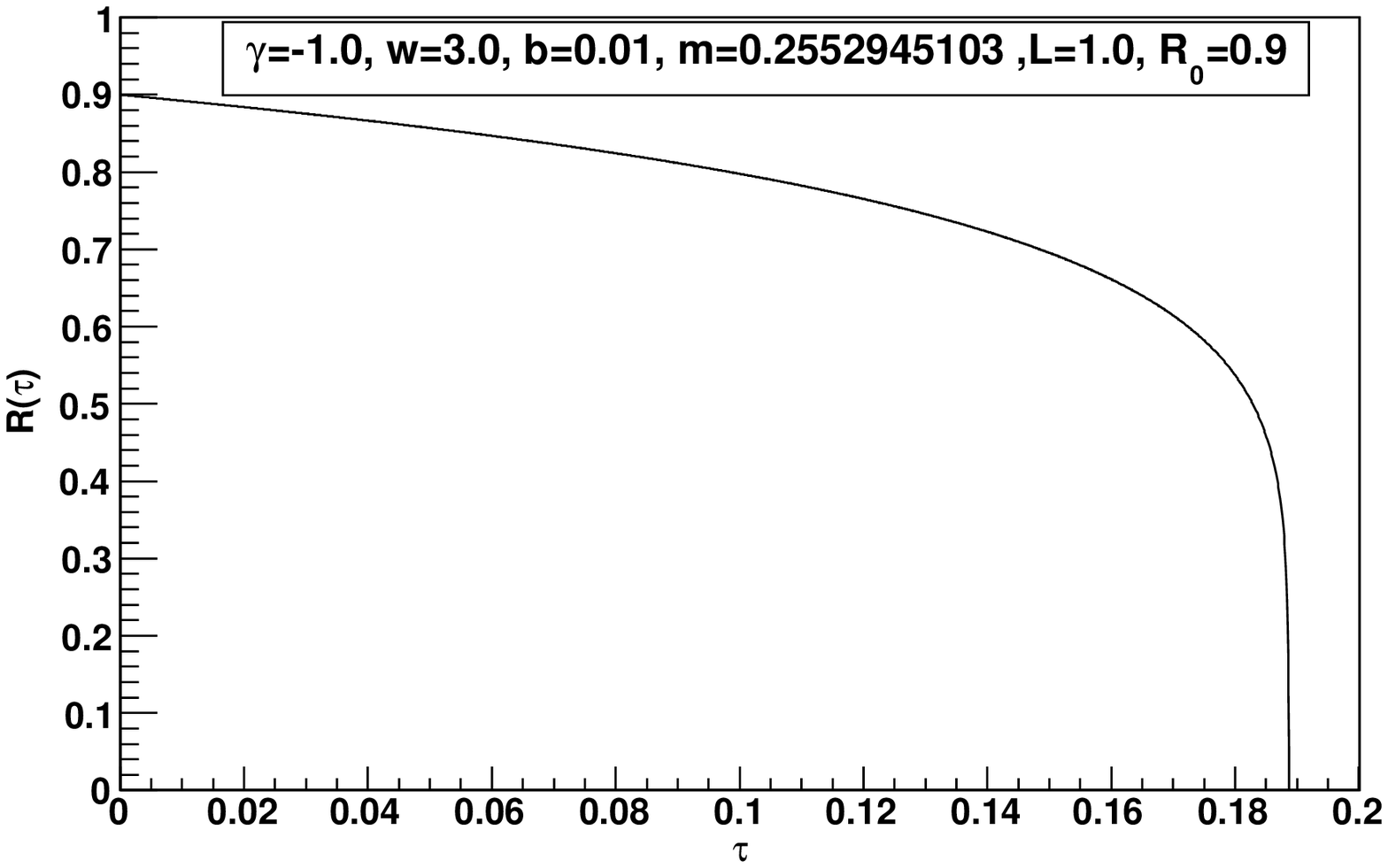,width=3.3truein,height=3.0truein}
\hskip .5in} \caption{These figures represent the dynamical evolution of shell to a 
naked singularity for the potential given by the figure \ref{fig5},
assuming $R_0=R(0)=0.9$.}
\label{fig6}
\end{figure}

An interesting result that we can deduce from Table III is that we can have
black hole and stable gravastar formation even with an interior and a shell 
constituted of dark and repulsive dark energy (cases H and I).
Still more interesting
is the case G, represented by figures \ref{fig1}, \ref{fig3} and \ref{fig5} 
where for small radius of the shell we
have no formation of a black hole (for $\Lambda=0$) and a naked singularity
(for  $\Lambda > 0$). This is the first time in the literature that a naked
singularity emerges from a gravastar model.  Besides, the figures \ref{fig2}, 
\ref{fig4} and \ref{fig6} give us examples of the dynamical evolution of a 
gravastar to a naked singularity.

Finally, the opposite final fates, because of the absence or the presence 
of a positive  cosmological constant,
reinforces the hypothesis proposed in \cite{nakedpressure} 
that can exist a connection between naked singularities and some kind of 
weakness of the gravitational field, compared to that associated to black holes.

\begin{table}
\caption{\label{tab:tablea} This table show the calculated horizons using 
the equations \ref{rbh} and \ref{rc}. The symbol $i=\sqrt{-1}$ denotes the imaginary constant. See figure \ref{fig1}.}
\begin{ruledtabular}
\begin{tabular}{cccc}
$m$ & $L_e$ & Cosmological Horizon  & Black Hole Horizon \\
\hline
0.2097639045 & 0.8 & 0.4795669330+0.2235024144 $i$ & 0.4795669330-0.2235024144 $i$ \\
	     & $L_e^*$ & & \\
             & 1.2 & 0.8577020327 & 0.5136246691 \\
             & 1.3104314 & 0.9974942529 & 0.4866382631 \\
             & 1.5 & 1.213250267 & 0.4638970952 \\
             & 10 & 9.783239233 & 0.4202701210 \\
             & $\infty$ & $\infty$ & 2m=0.4195278090 \\
\end{tabular}
\end{ruledtabular}
\end{table}                                                                     

\begin{table}
\caption{\label{tab:tableb} This table show the calculated horizons using 
the equations \ref{rbh} and \ref{rc}. The symbol $i=\sqrt{-1}$ denotes the imaginary constant. See figure \ref{fig3}.}
\begin{ruledtabular}
\begin{tabular}{cccc}
$m$ & $L_e$ & Cosmological Horizon  & Black Hole Horizon \\
\hline
0.3775 & 1.5 & 0.8943821768+0.3869863502 $i$ & 0.8943821767-0.3869863502 $i$ \\
	     & $L_e^*$ & & \\
       & 1.97 & 1.1976796420 & 1.0759952780 \\
       & 2.0053430150 & 1.2948106450 & 1.0151227670 \\
       & 3 & 2.5081539280 & 0.8151917235 \\
       & 20 & 19.61123828 & 0.7560805420 \\
       & $\infty$ & $\infty$ & 2m=0.7550 \\
\end{tabular}
\end{ruledtabular}
\end{table}                                                                     

\begin{table}
\caption{\label{tab:tablec} This table show the calculated horizons using 
the equations \ref{rbh} and \ref{rc}. The symbol $i=\sqrt{-1}$ denotes the imaginary constant. See figure \ref{fig5}.}
\begin{ruledtabular}
\begin{tabular}{cccc}
$m$ & $L_e$ & Cosmological Horizon  & Black Hole Horizon \\
\hline
0.2552945103 & 1.0 & 0.5973628039+0.2655691220 $i$ & 0.5973628039 - 0.2655691220 $i$ \\
	     & $L_e^*$ & & \\
             & 1.36 & 0.8837041274 & 0.6823811248 \\
             & 1.430886 & 1.002124260 & 0.6365867478 \\
             & 1.6 & 1.220128096 & 0.5913790740 \\
             & 12 & 11.73606139 & 0.5115184600 \\
             & $\infty$ & $\infty$ & 2m=0.5105890206 \\
\end{tabular}
\end{ruledtabular}
\end{table}

\begin{acknowledgments}
The financial assistance from FAPERJ/UERJ (MFAdaS) are gratefully acknowledged.
The authors (RC, MFAdaS, JFVR) acknowledges the financial support from FAPERJ (no. E-26/171.754/2000,
E-26/171.533/2002, E-26/170.951/2006, E-26/110.432/2009 and E26/111.714/2010). The authors (RC,
MFAdaS and JFVdR) also acknowledge the financial support from Conselho Nacional de Desenvolvimento Cient\'ifico e
Tecnol\'ogico - CNPq - Brazil (no. 450572/2009-9, 301973/2009-1 and 477268/2010-2). The author (MFAdaS)
also acknowledges the financial support from Financiadora de Estudos e Projetos - FINEP - Brazil
(Ref. 2399/03). 
\end{acknowledgments}


\begin{thebibliography}{88}
\bibitem{grava}  D. Horvat and S. Ilijic, arXiv:0707.1636;
                 P. Marecki, arXiv:gr-qc/0612178; 
		 F.S.N. Lobo, Phys. Rev. D{\bf 75},  024023 (2007); arXiv:gr-qc/0612030;
		   Class. Quantum Grav. {\bf 23},  1525 (2006);
		 F.S.N. Lobo, Aaron V. B. Arellano, {\em ibid.}, {\bf 24}, 1069   (2007);
		 T. Faber,  arXiv:gr-qc/0607029;
		 C. Cattoen, arXiv:gr-qc/0606011;
		 O.B. Zaslavskii, Phys. Lett. B{\bf 634},  111 (2006); 
		 C. Cattoen, T. Faber, and M. Visser, Class.  Quantum  Grav. {\bf 22},  4189 (2005).
\bibitem{DEs}   
   E.J. Copeland, M. Sami and S. Tsujikawa, Int. J. Mod. Phys. D{\bf 15}, 1753 (2006); 
   T. Padmanabhan,  arXiv:07052533.
\bibitem{BN07}  A.E. Broderick and R. Narayan,  Class.  Quantum  Grav. {\bf 24},  659  (2007) 
[arXiv:gr-qc/0701154].
\bibitem{MM01} P.O. Mazur and E. Mottola, ''{\em   Gravitational Condensate Stars: An Alternative to 
               Black Holes},'' arXiv:gr-qc/0109035;  Proc. Nat. Acad. Sci. {\bf 101},  9545 
	       (2004) [arXiv:gr-qc/0407075].
\bibitem{irina} Dymnikova I and Galaktionov E ,"{\em Vacuum Dark Fluid}",Physics Letters B {\bf 645},358 (2007) and references herein.
\bibitem{VW04} M. Visser and D.L. Wiltshire, Class. Quantum Grav. {\bf 21}, 1135  (2004)[arXiv:gr-qc/0310107].
\bibitem{Carter05} B.M.N. Carter,  Class. Quantum Grav. {\bf 22}, 4551  (2005)  [arXiv:gr-qc/0509087].
\bibitem{DeB06}  A. DeBenedictis, {\em et al}, Class.  Quantum  Grav. {\bf 23}, 2303   (2006) [arXiv:gr-qc/0511097].
\bibitem{CR07} C.B.M.H. Chirenti and L. Rezzolla,  arXiv:0706.1513.
\bibitem{JCAP} P. Rocha, A.Y. Miguelote, R. Chan, M.F. da Silva, N.O. Santos,and A. Wang,
    "{\em Bounded excursion stable gravastars and black holes}," J.
    Cosmol. Astropart. Phys. {\bf 6}, 25 (2008) [arXiv:gr-qc/08034200].
\bibitem{JCAP1} P. Rocha, R. Chan, M.F. da Silva and A. Wang,
    "{\em Stable and "Bounded Excursion" Gravastars, and Black Holes in Einstein's Theory of Gravity}," 
     J. Cosmol. Astropart. Phys. {\bf 11}, 10 (2008) [arXiv:gr-qc/08094879].
\bibitem{JCAP2} R. Chan, M.F. da Silva, P. Rocha and A. Wang,
    "{\em Stable Gravastars with Anisotropic Dark Energy}",
     J. Cosmol. Astropart. Phys. {\bf 3}, 10 (2009) [arXiv:gr-qc/08124924].
\bibitem{JCAP3} R. Chan, M.F.A. da Silva and P. Rocha,
    "{\em How the cosmological constant affects gravastar formation},"
     J. Cosmol. Astropart. Phys. {\bf 12}, 17 (2009) [arXiv:gr-qc/09102054].
\bibitem{JCAP4} R. Chan and M.F.A. da Silva,
    "{\em How the charge can affect the formation of gravastars},"
     J. Cosmol. Astropart. Phys. {\bf 7}, 29 (2010) [arXiv:gr-qc/10053703].
\bibitem{Chan} R. Chan, M.F.A. da Silva, J.F. Villas da Rocha,
 "{\em Star Models with Dark Energy}" (2008) [arXiv:gr-qc/08033064]
\bibitem{Paramos} Bertolami, O., P\'aramos, J., Phys. Rev. D {\bf 72}, 123512 
(2005) [arXiv:astro-ph/0509547]
\bibitem{Lobo07} Lobo, F. (2007) [arXiv:gr-qc/0611083].
\bibitem{CattoenVisser05} Cattoen, C., Faber, T. and Visser, M. Class. Quantum Grav. {\bf 22}
4189 (2005).
\bibitem{singleton}  Dzhunushaliev V, Folomeev V , Myrzakulov R and Singleton D,"{\em Non-singular solutions to Einstein-Klein-Gordon equations with a phantom scalar field}", Journal of High Energy Physics {\bf 7}, 94 (2008), [arXiv:gr-qc/arXiv:0805.3211].
\bibitem{Lobo} Lobo, F., Class. Quant. Grav. {\bf 23}, 1525 (2006).
\bibitem{Lake} Lake, K., Phys. Rev. D {\bf 19}, 2847 (1979).
\bibitem{HE73}  S.W. Hawking and G.F.R. Ellis, {\em The large scale structure of space-time}
 (Cambridge University Press, Cambridge, 1973).
\bibitem{Chan08} Chan, R., da Silva, M.F.A., Villas da Rocha, J.F., MPLA  {\bf 24 }, 1137, (2009). 
\bibitem{Shankaranarayanan} S. Shankaranarayanan, Phys. Rev. D. {\bf 67}, 084026 (2003).
\bibitem{nakedpressure} C. F. C. Brandt, Chan, R., da Silva, M.F.A., Villas da Rocha, J.F., IJMPD {\bf 19}, 317, (2010). 

\end{thebibliography}
\end{document}